\documentclass[final,5p,times,twocolumn]{elsarticle}

\usepackage{amssymb}
\usepackage{lineno,hyperref}
\usepackage{graphicx}	% Including figure files
\usepackage{amsmath}	% Advanced maths commands
\usepackage{amssymb}	% Extra maths symbols
\usepackage{subfig}
\usepackage{cleveref}
\usepackage{nameref}
\journal{Astroparticle Physics}

\begin{document}

\begin{frontmatter}

\title{A Two Population Electron Synchrotron Model For Knots of Extended Jets}
\author[label1]{Samaresh Mondal}

\ead{smondal@camk.edu.pl}
\author[label2]{Nayantara Gupta}

\address[label1]{Nicolaus Copernicus Astronomical Center, Polish Academy of Sciences, ul. Bartycka 18, 00-716 Warsaw, Poland}

\address[label2]{Raman Research Institute,

    C.V.Raman Avenue, Sadashivanagar, Bangalore 560080, India}

\begin{abstract}
Radio observations by ALMA and upper limits on gamma ray flux by Fermi LAT have ruled out inverse Compton scattering of  Cosmic Microwave Background radiation by relativistic electrons (IC/CMB) as the origin of X-ray emission from extended jets of six quasars 3C 273, PKS 0637-752, PKS 1136-135, PKS 1229-021, PKS 1354+195, and PKS 2209+080. 
Here we consider two populations of accelerated electrons in each knot of the extended jets to explain the radio to optical and the X-ray emission by synchrotron cooling of relativistic electrons. In all cases the jet power required is lower than the Eddington's luminosity and the observed knot emissions are well explained in this scenario.
\end{abstract}

\end{frontmatter}

\section{Introduction}
Several extended jets of distant quasars have been observed in radio frequencies by Atacama Large Millimeter/submillimeter Array (ALMA),  in optical frequencies by Hubble Space Telescope (HST), in X-rays by Chandra X-ray observatory and in gamma rays by Fermi LAT.  In most cases we only have upper limits on gamma ray fluxes. The multi-wavelength emission  from kilo-parsec scale jets is most often explained by synchrotron emission of relativistic electrons in radio and optical frequencies,  and inverse Compton scattering of Cosmic Microwave Background (CMB) radiation by relativistic electrons (IC/CMB) in X-ray frequencies \citep{Tavecchio et al.(2000),Celotti et al.(2001), Sambruna et al.(2008)}. 
The X-ray emission observed by {\it Chandra} from the 100 kilo-parsec scale jet of  PKS 0637-752 has been explained  earlier by IC/CMB model  by \citet{Tavecchio et al.(2000)}   assuming equipartition between magnetic field energy density and electron energy density and Doppler factor $\delta \sim 10$. The jet power required in their model is $10^{48}$ erg/sec. They noted that  synchrotron self Compton emission (SSC) requires much higher jet power to explain the X-ray emission from extended jet. In this case the Doppler factor  has to be very low as the minimum energy of the electrons is very high even for $\delta=1$.
{\it Chandra} data of the X-ray jet of 3C 273 was analysed and modelled by \citet{Sambruna et al.(2001)}. They discussed about the following scenarios to explain the X-ray emission
(i) synchrotron emission from a second more energetic population of electrons following the discussions by \citet{Roser et al.(2000)} (ii) SSC emission and (iii) IC/CMB emission.
 
If the second population of electrons is co-spatial with the first population then their acceleration has to happen more recently. With their limited data they could not constrain many of the parameters in this scenario. \citet{Uchiyama et al.(2006)} studied the multi-wavelength emission from the knots of the jet of 3C 273. They discussed that in the spectral energy distributions (SEDs) of the inner knots the low energy component extending from radio to infrared frequency and the high energy component extending from optical to X-ray frequency could be from synchrotron emission of two populations of electrons. Similar polarisations observed in radio and optical frequencies could be an evidence for synchrotron emission in optical frequencies, which also implies synchrotron origin of X-ray emission.
The two populations of electrons having different cut-off energies may be produced due to different acceleration rates in the knots. In section 4.3 of their paper the plausible scenarios for the generation of two populations of electrons have been discussed.

\citet{Sambruna et al.(2008)} have modelled the multi-wavelength data from the knots of S5 2007+777 with IC/CMB model assuming equipartition in energy between electrons and magnetic field. A high value of Doppler factor $\delta=13$ is needed in this case, which implies Mpc scale deprojected extension of the jet. If the energy in electrons is increased then the Doppler factor can be reduced. However this would lead to the production of higher flux of X-rays produced in IC/CMB. They also discussed the alternative scenario where the X-ray emission could be from a second population of more energetic electrons. The second population could be in the same region as the first population or in a different region as suggested by \citet{Jester et al.(2006)} for 3C 273. 
\par
\citet{Cara et al.(2013)} studied the extended jet emission of PKS 1136-135. They noted that several knots are highly polarised in optical emission. In some cases the optical and X-ray emission is from a single population of electrons. They discussed that in the IC/CMB model the minimum energy of the electrons has to be close to their rest mass and jet has to be highly beamed with Doppler factor $\delta \geq 20$. They suggested there could be a second population of electrons emitting optical and X-ray photons by synchrotron emission.
Better angular resolution in X-ray images is needed to understand whether the second population is co-spatial with the first population of electrons or they are in different regions.

\par
Although for some of the newly discovered extended jets the X-ray emission could be explained satisfactorily  with IC/CMB model \citep{Simionescu et al.(2016), Zacharias and Wagner(2016)}
in many other cases the gamma ray flux expected from IC/CMB model exceeds the upper limits from Fermi LAT  \citep{Meyer and Georganopoulos(2014), Meyer et al.(2015), Meyer et al.(2017)}, thus ruling out this model. IC/CMB model of X-ray emission has been ruled out for the extended jets of 3C 273 \citep{Meyer and Georganopoulos(2014)}, PKS 0637-752 \citep{Meyer et al.(2015), Meyer et al.(2017)} and more recently for the extended jets of some other quasars PKS 1136-135, PKS 1229-021, PKS 1354+195, and PKS 2209+080  \citep{Breiding et al.(2017)}.
 \par
The problems with IC/CMB model include high luminosity required to explain X-ray data,  high Lorentz factor required to explain jet emission which violates observed limit on Lorentz factor, offsets in peak brightness between radio, optical and X-ray  frequencies. Moreover polarisation measurements in UV frequency show excess polarisation  which is unexpected in IC/CMB model.
  
 The alternative scenario of proton synchrotron emission could be possible as protons lose energy very slowly by synchrotron emission  and traverse long distances before cooling down.
Proton synchrotron model has been applied to the extended jet of 3C 273 and PKS 0637-752 in earlier works \citep{Kundu and Gupta(2014),Bhattacharyya and Gupta(2016),Basumallick and Gupta(2017)}. 
The magnetic field needed to explain the X-ray emission in the proton synchrotron model is  of mG order for 3C 273 and PKS 0637-752.
   Most of the jet luminosity is due to this strong magnetic field.  
 The Lorentz factor of the emission region could be close to 3 in the proton synchrotron model, which is physically plausible as a kilo-parsec scale jet is expected to move much slower than a parsec scale jet. 
 But for the other extended jets proton synchrotron model requires Super-Eddington's luminosity to explain the X-ray emission. 
 \par
In this work we do a systematic analysis of synchrotron emission from two populations of electrons within a region of kilo-parsec scale radius to explain the radio to X-ray emission from knots of six sources. Although qualitatively this model has been discussed in earlier work, a quantative study including the latest observational data has not been done so far. We have assumed that the two populations are co-spatial.
 One population of injected electrons is having energy in the range of  MeV to GeV which would explain the radio to infrared data, another population having TeV energy would explain the optical to X-ray data. 
 The relativistic electrons lose energy by emitting synchrotron  and inverse Compton (IC) emission. The propagated electron spectra have been calculated with the publicly available time dependent code GAMERA \footnote{http://joachimhahn.github.io/GAMERA/docs}. The estimated radiations are compared with the observed data to determine the parameter values. The Tables with the parameters values are given in the APPENDIX.
 Our results are discussed in section 3.  
\section{Modelling of Spectral Energy Distributions}
\label{section_2}
The GAMERA code 
solves the transport equation of particles, in our case electrons, to get the time evolved particle spectrum and also calculates the radiation spectra.
 
The transport equation for the electron spectrum is
\begin{equation}\label{8}
\frac{\partial N(E,t)}{\partial t}=Q(E,t)-\frac{\partial}{\partial E}\Big(b(E,t) N(E,t)\Big)
\end{equation}
The injection spectrum is $Q(E,t)$ and the propagated spectrum is $N(E,t)$.
We have included the synchrotron loss of electrons and IC/CMB interactions in the transport equation by the term $b(E,t)$. This code uses the full Klein-Nishina cross-section for
IC scattering from \citet{Blumenthal and Gould(1970)} and self-consistently calculates the synchrotron and IC photon flux radiated by the relativistic electrons.
Diffusion loss is assumed to be negligible compared to the radiative losses of electrons.
The electron spectrum is evolved for $10^4$ years, which is assumed to be the age of the extended jet, and the synchrotron and IC spectra are calculated subsequently. 
 Power law electron spectrum with spectral index ($\alpha$) between 2 to 3 is injected into the spherical region of 1 kilo-parsec radius with minimum and maximum energies of electrons $E_{min}$ and $E_{max}$ respectively. The spectral index and minimum, maximum energies are adjusted to fit the data.

The jet power is calculated using this expression
\begin{equation}
P_{jet}=\pi R^2 \Gamma^2 c (u'_B+u'_e)
\end{equation}
where $u'_B$, $u'_e$ are the energy densities in magnetic field, electrons  respectively.  The jet power in magnetic field is denoted by $P_B$. In our case $u'_B >> u'_e$, so $P_{jet}\simeq P_{B}$.  This is necessary in our model to reduce the IC/CMB contribution. We note that in earlier studies authors often assumed equipartition in energy between magnetic field and electrons,
 to constrain the parameter values. This would give very high X-ray and gamma ray flux. We try to reduce the jet power by optimising the magnetic field $B$. We have calculated the total jet power $P_{total}$ after adding the jet power for all the knots and displayed in the Tables in APPENDIX.
$R$ is the radius of the spherical emission region moving with Lorentz factor $\Gamma$. 
 We have taken the emission regions of kilo-parsec scale radius motivated by the radio and X-ray observations of extended jets. It is important to note that Lorentz factor is expected to be low as extended jets move much slower than parsec scale jets \citep{Breiding et al.(2017)}. The Lorentz factor and the viewing angle are used to calculate the Doppler factor.
 If the Doppler factor is high we get more luminosity in IC/CMB emission in the observer's frame as it is boosted by a factor of $\delta^6/\Gamma^2$ \citep{Georganopoulos et al.(2006)}. We have chosen low values of Lorentz factor and high values of viewing angle $\theta_{obs}$, so that the values of the Doppler factor in our model are not high, and thus the luminosity in IC/CMB emission is low.
 Moreover,  the Doppler factor cannot be too low as this would require very high jet power to explain the multi-wavelength data with synchrotron emission of two 
 populations of electrons.

\paragraph{\bf S5 2007+777}
The BL Lac object S5 2007+777 at a redshift  of 0.342 having hybrid FR I and FR II radio morphology is classified as HYMOR object. 
The isotropic bolometric luminosity of S5 2007+777 is $5.12\times10^{45}$ erg/sec \citep{Nemmen et al.(2012)}.
The extended jet of S5 2007+777 has five knots $K_{3.6}$, $K_{5.2}$, $K_{8.5}$, $K_{11.1}$ and $K_{15.7}$. 
The distances of the knots from the core are measured in arcsecs in the 1.49 GHz image of VLA.
The radio data from knot $K_{8.5}$ is fitted with electron synchrotron and the X-ray data is fitted by two ways (i) IC/CMB (ii) synchrotron from a second population of electrons \citep{Sambruna et al.(2008)}. They assumed the jet to be strongly beamed with Doppler factor $\delta$=13. The superluminal feature found in VLBI observation constrains the jet to be aligned within $\thicksim$ $24^o$ to our line of sight. Which implies the deprojected length of the jet to be 150 kpc. 
The jet luminosity required to fit the data in the IC/CMB model is of the order of $10^{46}$ erg/sec.
In our study, we have used viewing angle $\theta_{obs}$=$20^o$ and Lorentz factor $\Gamma$ in the range of 2.4 to 2.8. The other parameters  values have been listed in Table 1. 
We have modelled the emission from each of the knots separately assuming the two populations of electrons in each knot are within spherical blobs of radius 1 kpc.
 All the parameter values of our model for the knots are given Table 1 and Table 2. Fig.1. shows the data points from knots fitted with electron synchrotron emission.
The total jet power required in the knots to model the radio and X-ray emission with electron synchrotron model is of the order of $6\times 10^{45}$ erg/sec, which is comparable to its bolometric luminosity .

\paragraph{\bf PKS 1136-135}
The FR II  object PKS 1136-135  located at  redshift 0.556 has been observed by Hubble Space Telescope and Chandra X-ray observatory and {\bf modelled by \citep{Sambruna et al.(2006)}}. The extended jet of this source has seven knots observed in radio and X-ray frequencies. They used IC/CMB model to fit the X-ray emission from the outer knots and synchrotron emission 
from a second population of electrons for the inner knots. It is interesting to note that the radio flux increases but the X-ray flux decreases with the distance of the knots from the core. The nearest knot $\alpha$ has the least radio flux and the furthest knot HS has the maximum radio flux. It was suggested that plasma deceleration could be the cause of decrease in the X-ray to radio flux ratio.
    
 The IC/CMB and two population electron synchrotron  models were revisited later \citep{Cara et al.(2013)}. 
  They discussed that IC/CMB model is disfavoured due to the high polarization in the optical data from the knots. The gamma ray flux expected from IC/CMB exceeds the upper limits from Fermi LAT for the first three knots $\alpha$, A and B
  \citep{Breiding et al.(2017)}. They also noted fitting the UV data with IC/CMB model for knot $\alpha$ and A over predicts the X-ray flux from them. Moreover the UV spectrum is harder than that expected from the IC/CMB model.
  The Eddington luminosity of PKS 1136-135 is $3.45\times10^{46}$ erg/sec \citep{Liu et al.(2006)}.  The two population electron synchrotron model used in our work requires total luminosity $10^{46}$ erg/sec to explain the observed emission from the knots. The parameter values used in our modelling are given in Table 3 and Table 4. We have assumed low values of Lorentz factors and large values of viewing angles to lower the values of the Doppler factors. Fig.2. shows the data points from the seven knots fitted with our model.
  
 \paragraph{\bf PKS 1229-021}
 PKS 1229-021 located at a redshift of 1.045 has been observed by Chandra X-ray observatory \citep{Weisskopf et al.(2000)} and Hubble Space Telescope \citep{Tavecchio et al.(2007)}. VLA image at 8.4GHz frequency shows four knots A, B, C and D.
Atacama Large Millimeter/submillimeter Array (ALMA) and Chandra have resolved knot A in the extended jet of PKS 1229-021 but the emission from the knots B, C, D could not be resolved separately. More observational data from the
 separate knots would be helpful to model the spectral energy distribution from each knot separately. 
 The data points fitted in Fig.3. have been taken from \citep{Breiding et al.(2017)}. Eddington luminosity of PKS 1229-021 is $6.31\times10^{46}$erg/sec \citep{Liu et al.(2006)}. The  values of the parameters used in this work are given in Table 5 and Table 6. In this case also the total jet power required in our model is lower than the Eddington's luminosity of this source. 

\paragraph{\bf PKS 1354+195}
 PKS 1354+195 located at redshift 0.72 has a bright core. VLA and ALMA observations resolved two knots A and B in its extended jet. Due to the bright core it is difficult to resolve knot A in X-rays. 
  Its jet emission has been modelled by \citep{Sambruna et al.(2002)} and \citep{Sambruna et al.(2004)} with IC/CMB model. We have taken the observed X-ray flux at 1 KeV from \citep{Sambruna et al.(2004)} and other multi-wavelength observational data  from \citep{Breiding et al.(2017)}. The Eddington luminosity of this quasar is $2.93\times10^{47}erg/sec$ \citep{Liu et al.(2006)}. The multi-wavelength data and the fittings of the SEDs are shown in Fig.4.
  Table 7 and Table 8 display the values of the parameters used in our model.  Total jet power required in our model is more than 10 times less than the Eddington's luminosity of this source.

\paragraph{\bf PKS 2209+080}
 PKS 2209+080 located at redshift 0.485 has five knots A, B, C, D, E in its extended jet. The extended jet emission was modelled with IC/CMB earlier \citep{Jorstad and Marscher(2006)}. Only one X-ray data point was noted from knot E in this 
paper. Due to the lack of X-ray data from the knots we can not constrain the values of the model parameters.
 The radio and optical data are available for all the knots. From electron synchrotron modelling of these data we constrain the magnetic field. 
 Future observations with Chandra would be helpful for better understanding of knot emission.
  The mass of its central blackhole is yet unknown. Assuming it to be $10^{9}M_{\odot}$ the Eddington luminosity is expected to be $1.22\times10^{47}$erg/sec. Fig.5. shows the observed fluxes and the SEDs from our model.
The values of the parameters used in the present work are given in Table 9 and Table 10. The total jet power required in electron synchrotron model $4\times 10^{46}$ erg/sec is less than half of its Eddington's luminosity.

\paragraph{\bf PKS 0637-752}
 PKS 0637-752 located at a redshift of 0.651 has four knots in its extended jet.  
 IC/CMB model has been ruled out for the jet emission by ALMA data and Fermi LAT upper limits \citep{Meyer et al.(2015),Meyer et al.(2017)}. It has been shown earlier that the X-ray emission from the knots could be explained by proton synchrotron model \citep{Basumallick and Gupta(2017)}.
 The Eddington luminosity of PKS 0637-752 could be of the order of $10^{48}$ erg/sec \citep{Kusunose and Takahara(2017)}. 
 The values of the parameters used in our model are displayed in Table 11 and Table 12. The total jet power required is ten times lower than the Eddington's luminosity of this source.
 Fig.6. shows the observed and our calculated fluxes.
\section{Results and Discussions}
The SEDs fitted in our model to the data from the knots of the extended jets of six quasars are shown in Fig.1. to Fig.6. Table 1 to Table 12 have the parameter values used to fit the data. 
 The magnetic fields in the knots are varied in the range of 0.03 mG to 0.7 mG to fit the observed radio and X-ray data.
  \par
  Fig.1. shows the fit to the data points for the source S5 2007+777. The first population of injected electrons is having energy in the range of  10 MeV to 10 GeV and spectral index at injection is in the range of 2.3 to 3. The second population has energy at injection in the range of  10 to 100 TeV. The transport equation calculates the time evolved electron spectrum after including radiative losses by electrons. Our formalism does not include any escape or diffusion loss of electrons. 
  Synchrotron loss by lower energy electrons gives the radio and optical flux (dashed lines in figures) and the same from the high energy population gives the X-ray flux (solid lines). IC/CMB emission gives high energy gamma rays as shown in the figures.
  The high energy gamma rays will be attenuated due to interactions with the extragalactic background light (EBL). We have not included the EBL correction on high energy gamma rays in this work.
  
  \par
 Fig.2. shows the emissions  from the knots of  PKS 1136+135. We have many low energy data points from the knots of its extended jet. The radio fluxes from the knots increase with their distances from the core. But a similar variation in the X-ray fluxes from the knots has not been observed. In Table 3 the magnetic field varies in the range of  0.03 mG to 0.15 mG. The total jet power required in the knots ($10^{46}$ erg/sec) is lower than the Eddington luminosity $3.45\times 10^{46}$ erg/sec of this source.  
 The Lorentz factor is always assumed to be below 3 as constrained from proper motion study of the knots of 3C 273 \citep{Meyer et al.(2016)}.
 We do not have observational constraints on the viewing angles for the extended jets of PKS 1136+135, PKS 1229-021, PKS 1354+195, PKS 2209+080 and PKS 0637-752. A moderately high value of $20^o$ is assumed so that Doppler factor is not too high.
 \par
 
 Fig.3. shows the multi-wavelength emission from knot A and combined emission from knots B, C and D of  PKS 1229-021. In future with more observational data it would be possible to study the knot to knot variations of the emissions from B, C and D. The total jet power
  required $1.32\times10^{46}$ erg/sec as given in Table 6  is lower than its Eddington luminosity $6.31\times 10^{46}$ erg/sec. The magnetic field is assumed in the range of 0.1 to 0.2 mG. The size of the emission region is always assumed to be 1 kpc.
\par
The extended jet of PKS 1354+195 has two knots A and B.  Fig.4. shows the SEDs of these knots. The knot A nearer to the core is brighter in radio and X-ray frequencies than the knot B. We have adjusted the values of the magnetic field and the Doppler factor such that the IC/CMB emission is not too high.

 \par
 Fig.5. shows the data points and SEDs from the five knots of PKS 2209+080. In future with better  observational data in X-ray frequencies the modelling could be improved to constrain the parameter values. The nearest knot is brightest in radio. But the
 radio fluxes do not decrease systematically with the distances of the knots from the core. The parameter values are shown in Table 9 and Table 10. The magnetic fields in the knots vary in the range of 0.05 mG to 0.3 mG and the spectral indices of the injected electrons are in the range of 2.1 to 2.3. The total jet power required is less than $50\%$ of its Eddington's luminosity.
 
  \par
 Fig.6. shows the multi-wavelength emissions from the knots of PKS 0637-752. The radio and X-ray fluxes are the least from the nearest knot $WK_{5.7}$.  With more observational data in X-ray frequencies the modelling could be improved in future. 
 The total jet power required is lower than the Eddington's luminosity of this source. The parameter values are displayed in Table 11 and Table 12. For this source proton synchrotron model also can explain the X-ray emission from knots without exceeding the luminosity budget.
\par
In PKS 1136-135 for several knots a high degree of UV polarization has been observed which indicates synchrotron origin.
It has been discussed earlier (see section 4.2 of \citet{Breiding et al.(2017)} that the UV upturns in the SEDs of knots A-C of  PKS 2209+080, knots $\alpha$ and A in PKS 1136-135 and hardening of the optical-UV spectrum indicates a synchrotron origin from a second population of electrons.

We have fitted the UV upturns in the SEDs of knots $\alpha$ and A of PKS 1136-135, knots A-C of PKS 2209+080 and knot $\rm{WK_{5.7}}$ of PKS 0637-752 with synchrotron emission of the second population of electrons as shown in Fig.2., Fig.5. and Fig.6. respectively.
\par
Finally it is difficult to explain what kind of acceleration mechanism can produce two populations of electrons in two energy ranges. There could be different acceleration mechanisms working simultaneously to accelerate the electrons to different energy ranges. 

Following the discussion by \citet{Uchiyama et al.(2006)} the acceleration time scale $t_{acc}=\xi r_g/c$, where the gyroradius $r_g=E/(e B)$. The rate of accelerations is determined by the factor $\xi \geq 1$, which is usually assumed to be energy independent. The maximum  energy of the electrons could be different due to different values of the $\xi$ factor.
\par
Moreover the factor $\xi$ is approximately the ratio of energy densities in regular and turbulent components of magnetic field  for relativistic shocks in the theory of diffusive shock acceleration. Development of a different turbulent component of magnetic field near the shock may also lead to a different value of $\xi$. Another possibility could be two different shocks with different values of $\xi$ generating two populations of electrons.

Further investigation in this direction is necessary to understand the physics of knot emissions in extended jets.
\begin{figure*}
\centering
\includegraphics[width=.8\textwidth]{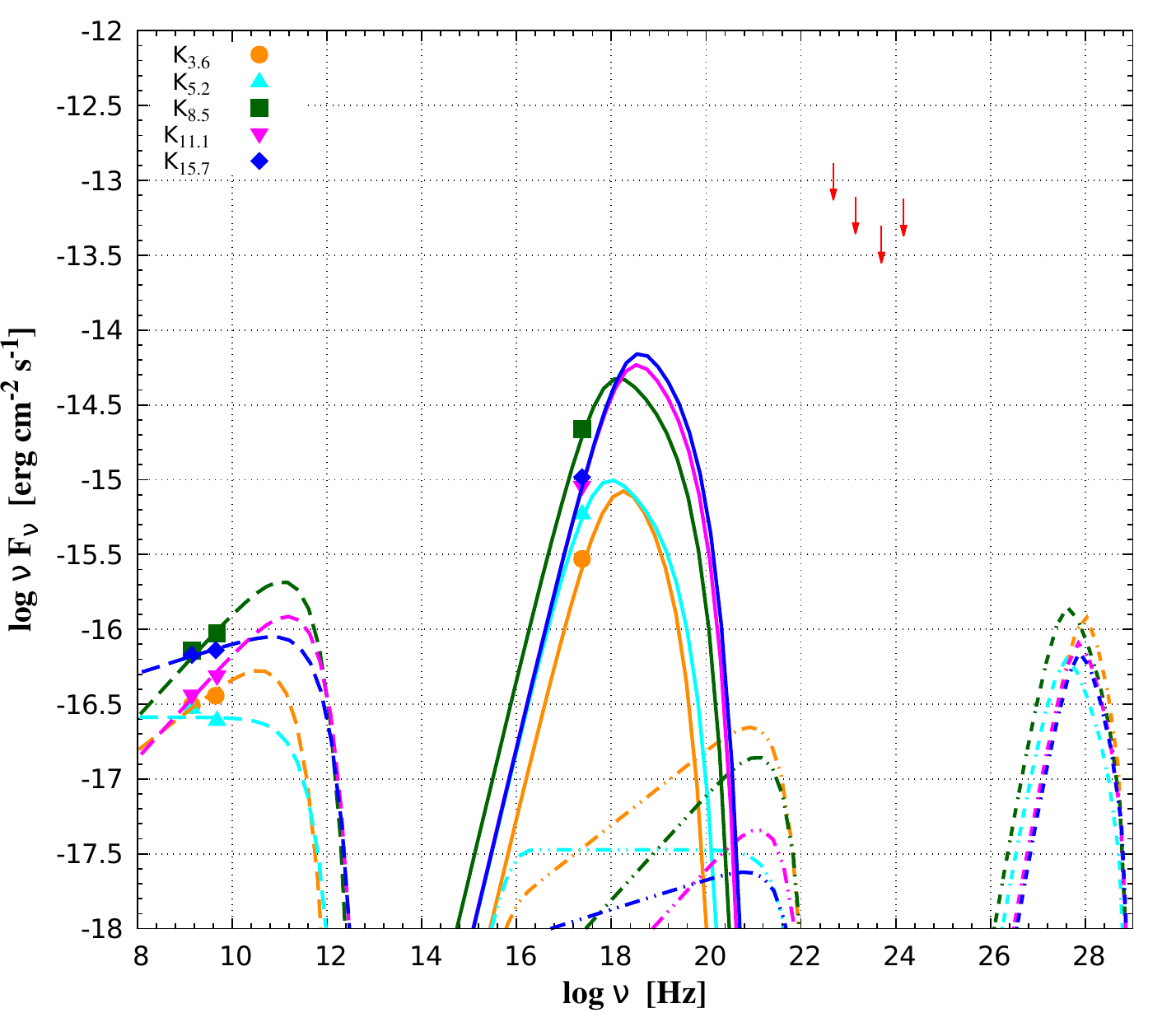}
\caption{\footnotesize{Multi-wavelength data of the knots in the extended jet of S5 2007+777 from \citep{Sambruna et al.(2008)}. The dashed lines represent electron synchrotron flux from the first population of electrons and the solid lines are for the second population. Dashed double dotted lines represent IC/CMB emission from first population of electrons and short dashed lines represent the same from the second population of electrons.
Different colours and types of points represent different knots. Fermi LAT upper limits are shown with downward red arrows.}}
\label{Figure-1}
\end{figure*}

\begin{figure*}
\centering
\includegraphics[width=.8\textwidth]{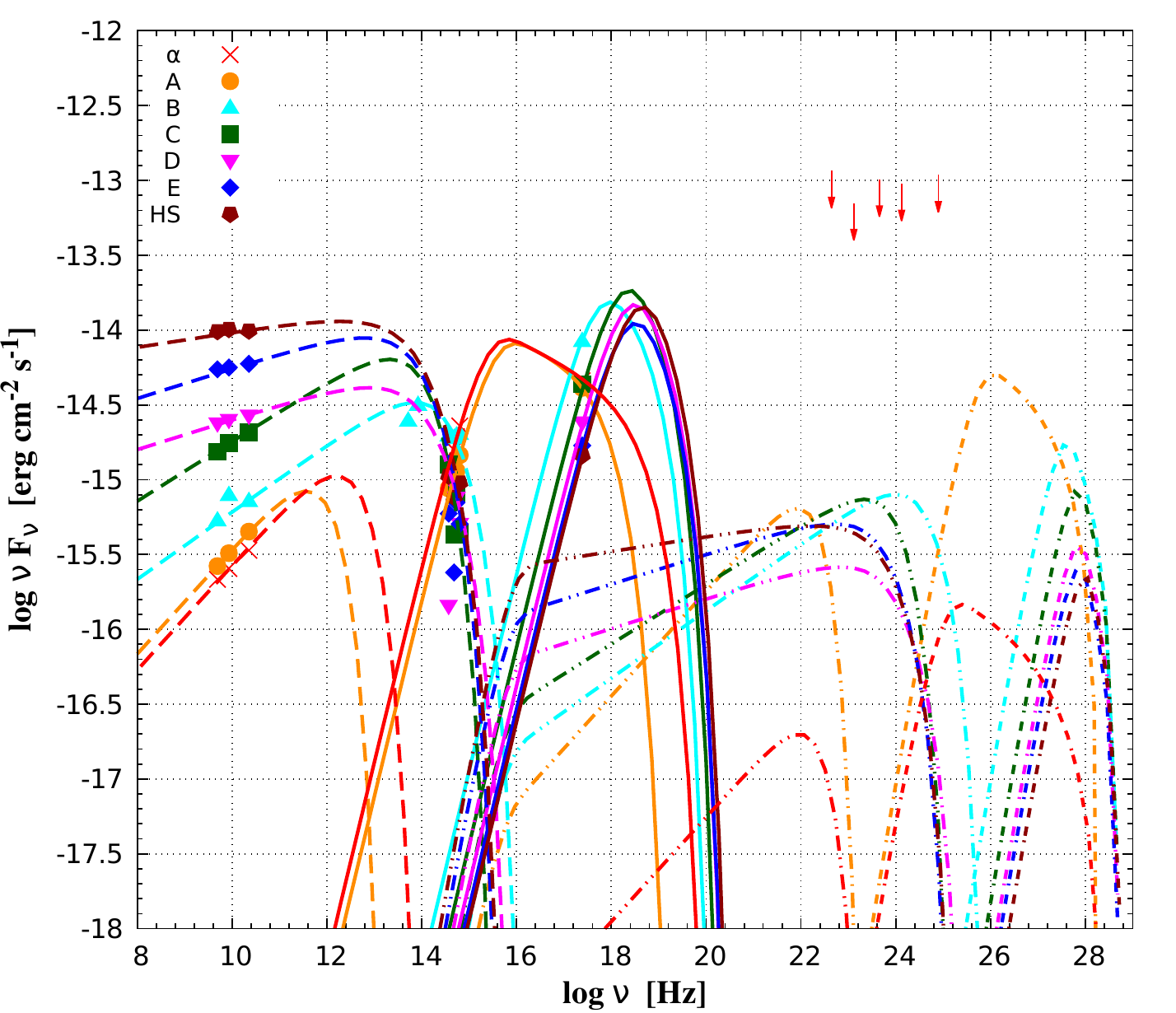}
\caption{\footnotesize{Multi-wavelength data of the knots in the extended jet of PKS 1136-135 from \citep{Sambruna et al.(2006)}. The data points for wavelengths 555nm, $5.8{\mu}m$, $3.6{\mu}m$ have been taken from \citep{Breiding et al.(2017)}. Different colours and types of points are showing the different knots. Red downward arrows represent Fermi LAT upper limits. The line styles are same as in Fig.1.
}}
\label{Figure-2}
\end{figure*}

\begin{figure*}
\centering
\includegraphics[width=.8\textwidth]{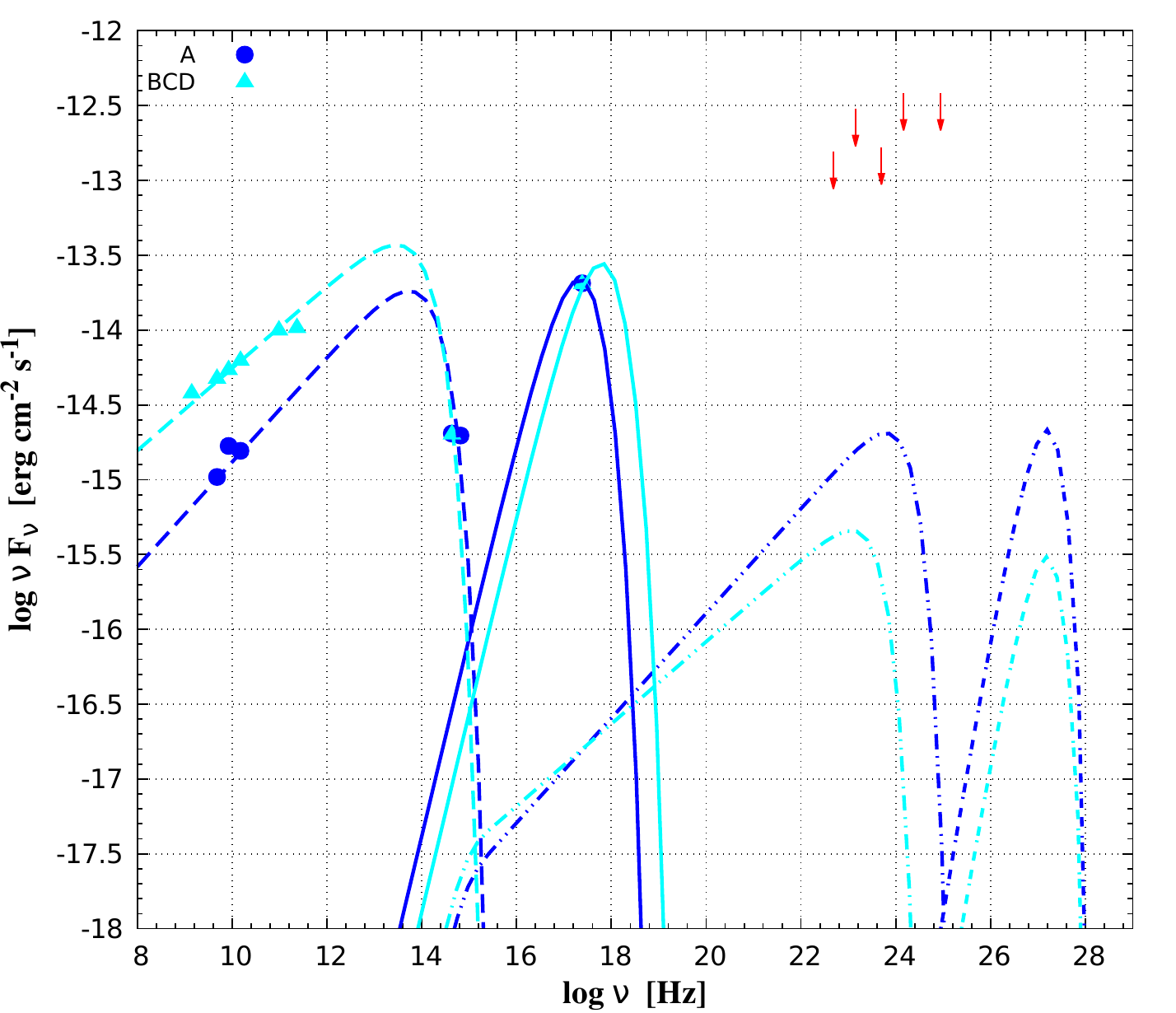}
\caption{\footnotesize{Multi-wavelength data of the knots in the extended jet of PKS 1229-021. Data points have been taken from \citep{Breiding et al.(2017)}. BCD represents combined flux from knots B, C, D as these knots could not be resolved separately.  Red downward arrow represents Fermi LAT upper limits. The line styles are same as in Fig.1.}}
\label{Figure-3}
\end{figure*}
\begin{figure*}
\centering
\includegraphics[width=.8\textwidth]{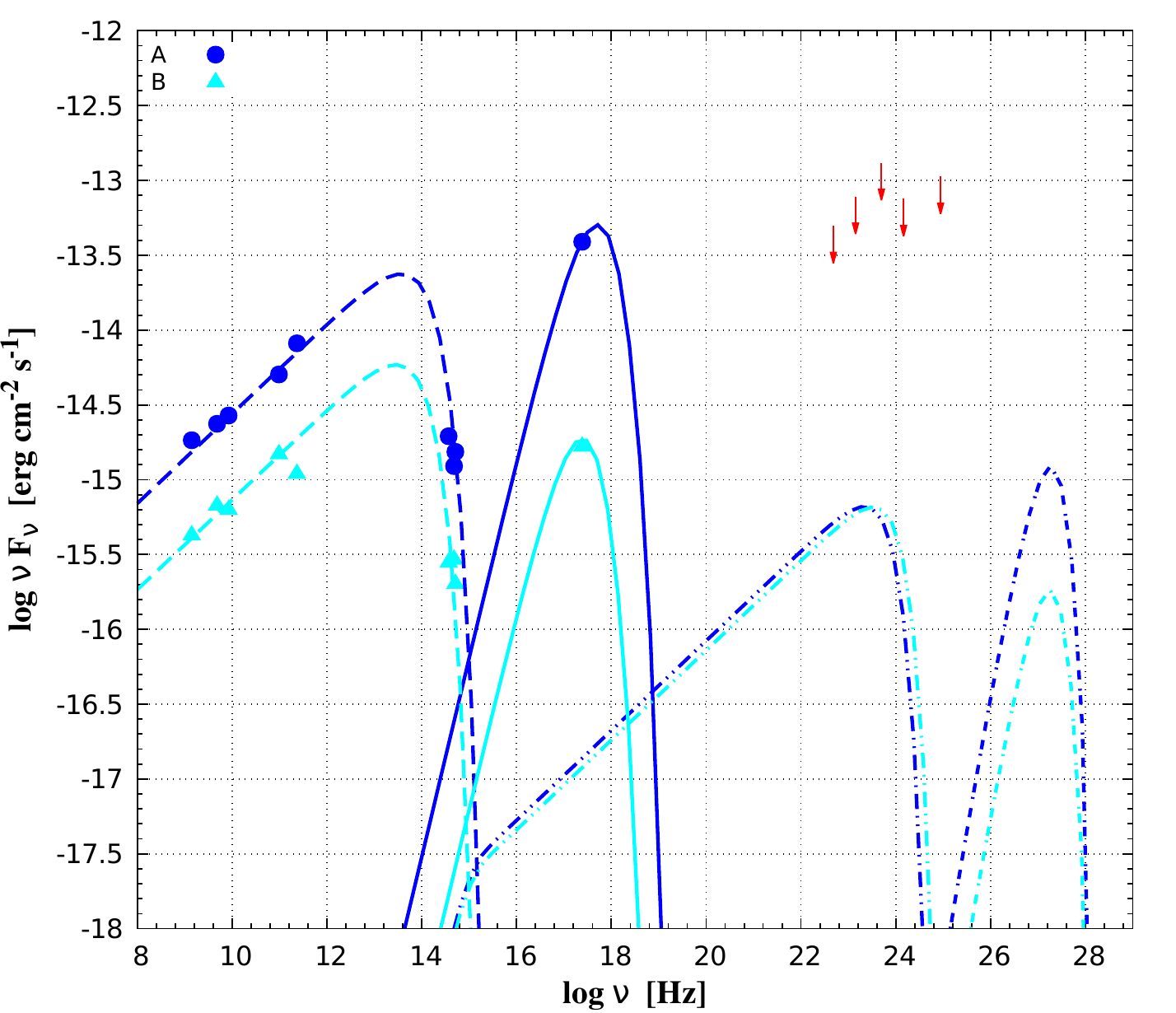}
\caption{\footnotesize{Multi-wavelength data from the knots of the extended jet of PKS 1354+195. Data points and Fermi LAT upper limits taken from \citep{Breiding et al.(2017)}. Line styles are same as in Fig.1..}}
\label{Figure-4}
\end{figure*}

\begin{figure*}
\centering
\includegraphics[width=.8\textwidth]{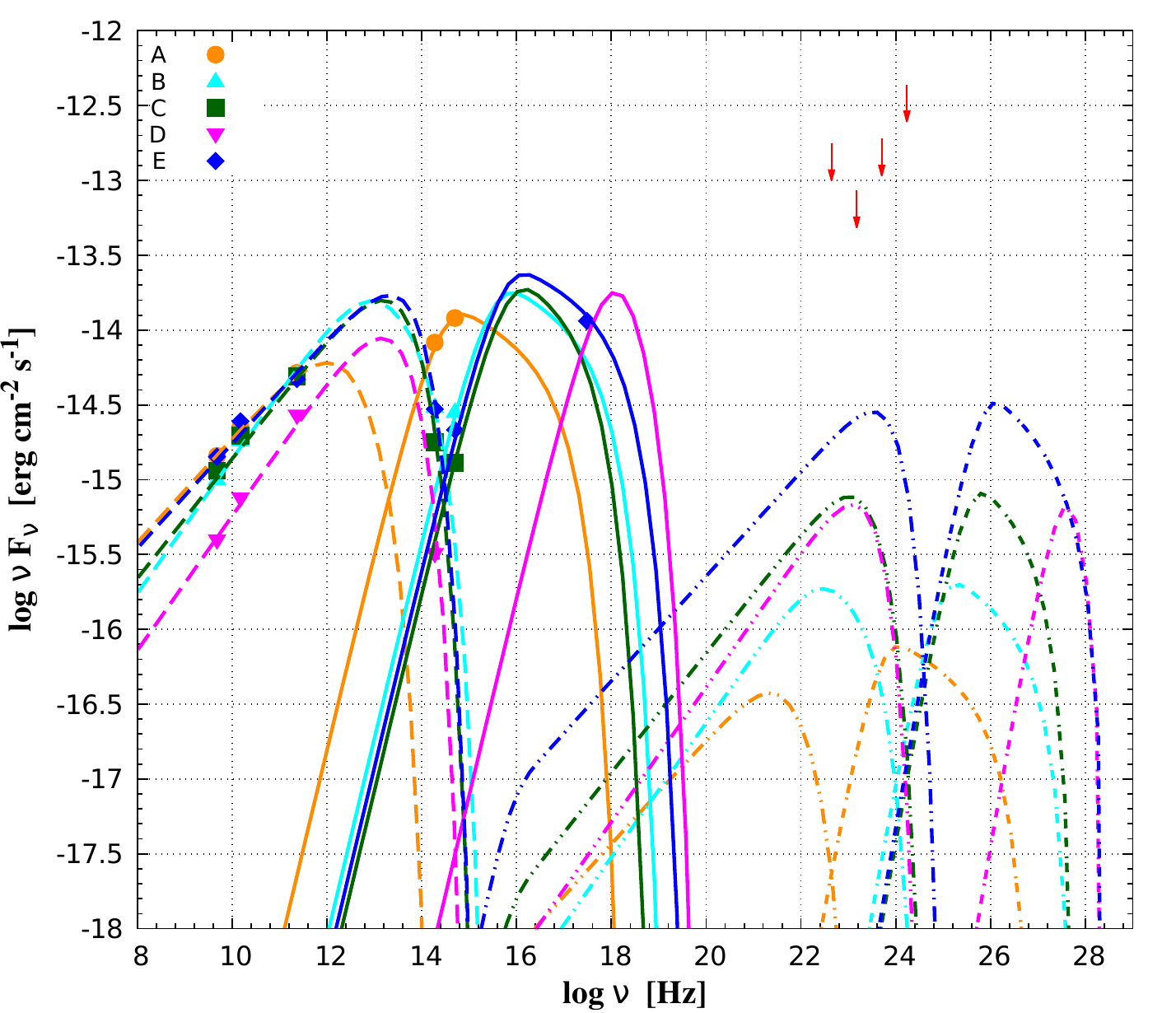}
\caption{\footnotesize{Multi-wavelength data from the five knots of the extended jet of PKS 2209+080. Data points, Fermi LAT upper limits are as given in \citep{Breiding et al.(2017)}.}}
\label{Figure-5}
\end{figure*}

\begin{figure*}
\centering
\includegraphics[width=.8\textwidth]{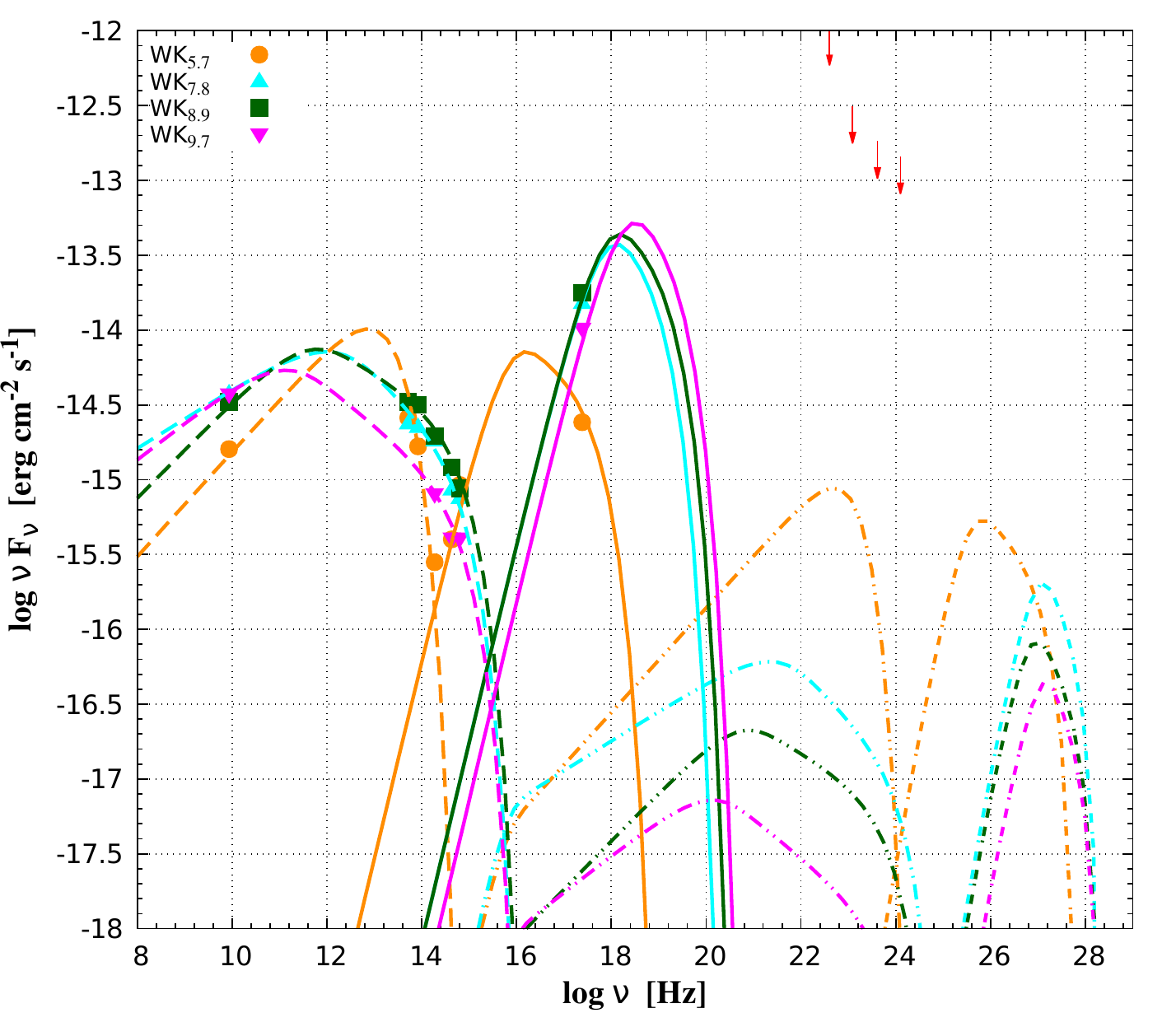}
\caption{\footnotesize{Multi-wavelength data from the knots of the extended jet of PKS 0637-752. Data point have been taken from \citep{Mehta et al.(2009)}, Fermi LAT upper limits from \citep{Meyer et al.(2015)} and line styles are same as before.}}
\label{Figure-6}
\end{figure*}
\section{Acknowledgment}
We thank the referee for constructive comments.

\section{APPENDIX}

\begin{table*}
\centering
\caption{\bf{Parameters used for modelling extended jet emission of S5 2007+777 with synchrotron emission from two populations of electrons. Second column displays the distances of the knots from the core, third column Lorentz factor, fourth column viewing angle, fifth column Doppler factor, sixth column magnetic field, seventh column type of emission, eighth and ninth columns minimum and maximum energies of electrons, and tenth column spectral index of injected electron spectrum.}}

\scalebox{1.0}{
\begin{tabular}{|c|c|c|c|c|c|c|c|c|c|} 

\hline

Knots & Dist(arcsec) & $\Gamma$ & $\theta_{obs}$ & $\delta$ & $B(mG)$ & Emission & $E_{min}(eV)$ & $E_{max}(eV)$ & $\alpha$\\ \hline

$K_{3.6}$ & 3.6 & 2.8 & $20^{\circ}$ & 5.92 & 0.03 & $1st\ population\ e^{-}synch$ & $10^7$ & $10^{10}$ & 2.5\\ \cline{7-10}
&&&&&& $2nd\ population\ e^{-}synch$ & $1.7\times10^{13}$ & $10^{14}$ & 2.3\\ \hline

$K_{5.2}$ & 5.1& 2.7 & $20^{\circ}$ & 6.14 & 0.05 & $1st\ population\ e^{-}synch$ & $10^7$ & $10^{10}$ & 3\\ \cline{7-10}
&&&&&& $2nd\ population\ e^{-}synch$ & $8.91\times10^{12}$ & $10^{14}$ & 2.3\\ \hline

$K_{8.5}$ & 8.5 & 2.6 & $20^{\circ}$ & 6.37 & 0.07 & $1st\ population\ e^{-}synch$ & $10^7$ & $10^{10}$ & 2.3\\ \cline{7-10}
&&&&&& $2nd\ population\ e^{-}synch$ & $8.91\times10^{12}$ & $10^{14}$ & 2.3\\ \hline

$K_{11.1}$ & 11 & 2.5 & $20^{\circ}$ & 6.63 & 0.09 & $1st\ population\ e^{-}synch$ & $10^7$ & $10^{10}$ & 2.3\\ \cline{7-10}
&&&&&& $2nd\ population\ e^{-}synch$ & $1.25\times10^{13}$ & $10^{14}$ & 2.3\\ \hline

$K_{15.7}$ &16 & 2.4 & $20^{\circ}$ & 6.9 & 0.1 & $1st\ population\ e^{-}synch$ & $10^7$ & $10^{10}$ & 2.4\\ \cline{7-10}
&&&&&& $2nd\ population\ e^{-}synch$ & $1.25\times10^{13}$ & $10^{14}$ & 2.3\\ \hline

\end{tabular}}
\vspace{0.2cm}
\label{Table-1}
\end{table*}

\begin{table*}

\centering
\caption{\textbf{Magnetic field energy densities of different knots and total jet power for object S5 2007+777}}

\scalebox{1.0}{
\begin{tabular}{|c|c|c|c|c|c|}\hline

Knots & $u'_{B}(erg/cm^3)$ & {$R$} (cm) & $P_{B}$(erg/sec) & $P_{total}$(erg/sec)\\ \hline

$K_{3.6}$ & $3.58\times10^{-11}$ & $3.086\times10^{21}$ & $2.52\times10^{44}$ & $5.93\times10^{45}$\\\cline{1-4}

$K_{5.2}$ & $9.95\times10^{-11}$ & $3.086\times10^{21}$ & $6.51\times10^{44}$ & \\\cline{1-4}

$K_{8.5}$ & $1.95\times10^{-10}$ & $3.086\times10^{21}$ & $1.18\times10^{45}$ & \\\cline{1-4}

$K_{11.1}$ & $3.22\times10^{-10}$ & $3.086\times10^{21}$ & $1.8\times10^{45}$ & \\\cline{1-4}

$K_{15.7}$ & $3.98\times10^{-10}$ & $3.086\times10^{21}$ & $2.05\times10^{45}$ & \\\hline

\end{tabular}}
\vspace{0.2cm}
\label{Table-2}
\end{table*}

\begin{table*}
\centering
\caption{\textbf{Parameters used for modelling multi-wavelength data from extended jet of PKS 1136-135 }}

\scalebox{1.0}{
\begin{tabular}{|c|c|c|c|c|c|c|c|c|c|} 

\hline

Knots & Dist(arcsec)&$\Gamma$ & $\theta_{obs}$ & $\delta$ & $B(mG)$ & Emission & $E_{min}(eV)$ & $E_{max}(eV)$ & $\alpha$\\ \hline

$\alpha$ & 2.7 & 2.2 & $20^{\circ}$& 7.53 & 0.15& $1st\ population\ e^{-}synch$ & $10^7$ & $3.16\times10^{10}$ & 2.3\\ \cline{7-10}
&&&&&& $2nd\ population\ e^{-}synch$ & $3.71\times10^{11}$ & $3.16\times10^{13}$ & 2.3\\ \hline

$A$& 4.6 & 2.8 & $20^{\circ}$& 5.92& 0.03 & $1st\ population\ e^{-}synch$ & $10^7$ & $3.16\times10^{10}$ & 2.3\\ \cline{7-10}
&&&&&& $2nd\ population\ e^{-}synch$ & $1.12\times10^{12}$ & $3.16\times10^{13}$ & 2.3\\ \hline

$B$ & 6.5 & 2.7 & $20^{\circ}$ & 6.14 & 0.05 & $1st\ population\ e^{-}synch$ & $10^7$ & $7.94\times10^{11}$ & 2.55\\ \cline{7-10}
&&&&&& $2nd\ population\ e^{-}synch$ & $10^{13}$ & $6.3\times10^{13}$ & 2.3\\ \hline

$C$ & 7.7 & 2.6 & $20^{\circ}$ & 6.37 & 0.07 & $1st\ population\ e^{-}synch$ & $10^7$ & $3.16\times10^{11}$ & 2.6\\ \cline{7-10}
&&&&&& $2nd\ population\ e^{-}synch$ & $1.14\times10^{13}$ & $6.3\times10^{13}$ & 2.3\\ \hline

$D$ & 8.6& 2.5 & $20^{\circ}$ & 6.63 & 0.09 & $1st\ population\ e^{-}synch$ & $10^7$ & $4.46\times10^{11}$ & 2.8\\ \cline{7-10}
&&&&&& $2nd\ population\ e^{-}synch$ & $1.41\times10^{13}$ & $6.3\times10^{13}$ & 2.3\\ \hline

$E$ & 9.3& 2.4 & $20^{\circ}$ & 6.9 & 0.095 & $1st\ population\ e^{-}synch$ & $10^7$ & $3.16\times10^{11}$ & 2.8\\ \cline{7-10}
&&&&&& $2nd\ population\ e^{-}synch$ & $1.41\times10^{13}$ & $6.3\times10^{13}$ & 2.3\\ \hline

$HS$ & 10.3 & 2.3 & $20^{\circ}$ & 7.2& 0.1 & $1st\ population\ e^{-}synch$ & $10^7$ & $3.16\times10^{11}$ & 2.9\\ \cline{7-10}
&&&&&& $2nd\ population\ e^{-}synch$ & $1.58\times10^{13}$ & $6.3\times10^{13}$ & 2.3\\ \hline

\end{tabular}}
\vspace{0.2cm}
\label{Table-3}
\end{table*}

\begin{table*}
\centering
\caption{\textbf{Magnetic field energy densities of different knots and total jet power for object PKS 1136-135}}

\scalebox{1.0}{
\begin{tabular}{|c|c|c|c|c|}\hline

Knots & $u'_{B}(erg/cm^3)$ & {$R$} (cm) & $P_{B}$(erg/sec) & $P_{total}$(erg/sec)\\ \hline

$\alpha$ & $8.95\times10^{-10}$ & $3.086\times10^{21}$ & $3.89\times10^{45}$ & $1.15\times10^{46}$\\\cline{1-4}

$A$ & $3.58\times10^{-11}$ & $3.086\times10^{21}$ & $2.52\times10^{44}$ & \\\cline{1-4}

$B$ & $9.95\times10^{-11}$ & $3.086\times10^{21}$ & $6.51\times10^{44}$ & \\\cline{1-4}

$C$ & $1.95\times10^{-10}$ & $3.086\times10^{21}$ & $1.18\times10^{45}$ & \\\cline{1-4}

$D$ & $3.22\times10^{-10}$ & $3.086\times10^{21}$ & $1.8\times10^{45}$ & \\\cline{1-4}

$E$ & $3.59\times10^{-10}$ & $3.086\times10^{21}$ & $1.85\times10^{45}$ & \\\cline{1-4}

$HS$ & $3.98\times10^{-10}$ & $3.086\times10^{21}$ & $1.88\times10^{45}$ & \\\hline

\end{tabular}}
\vspace{0.2cm}
\label{Table-4}
\end{table*}

\begin{table*}
\centering
\caption{\textbf{Parameters used for modelling multi-wavelength data from extended jet of PKS 1229-021}}

\scalebox{1.0}{
\begin{tabular}{|c|c|c|c|c|c|c|c|c|c|} 

\hline

Knots & Dist(arcsec) & $\Gamma$ & $\theta_{obs}$ & $\delta$ & $B(mG)$ & Emission & $E_{min}(eV)$ & $E_{max}(eV)$ & $\alpha$\\ \hline

$A$ & 0.7 & 2.7 & $20^{\circ}$ & 6.14 & 0.2 & $1st\ population\ e^{-}synch$ & $10^7$ & $5.01\times10^{11}$ & 2.3\\ \cline{7-10}
&&&&&& $2nd\ population\ e^{-}synch$ & $5.62\times^{12}$ & $3.16\times10^{13}$ & 2.3\\ \hline

$BCD$&1.9 & 2.8 & $20^{\circ}$& 5.92 & 0.1 & $1st\ pouplation\ e^{-}synch$ & $10^7$ & $3.16\times10^{11}$ & 2.45\\ \cline{7-10}
&&&&&& $2nd\ population\ e^{-}synch$ & $5.62\times10^{12}$ & $3.16\times10^{13}$ & 2.3\\ \hline

\end{tabular}}
\vspace{0.2cm}
\label{Table-5}
\end{table*}

\begin{table*}
\centering
\caption{\textbf{Magnetic field energy densities of different knots and total jet power for object PKS 1229-021}}

\scalebox{1.0}{
\begin{tabular}{|c|c|c|c|c|}\hline

Knots & $u'_{B}(erg/cm^3)$ & {$R$} {cm} & $P_{B}$(erg/sec) & $P_{total}$(erg/sec)\\ \hline

$A$ & $1.59\times10^{-9}$ & $3.086\times10^{21}$ & $1.04\times10^{46}$ & $1.32\times10^{46}$\\\cline{1-4}

$BCD$ & $3.98\times10^{-10}$ & $3.086\times10^{21}$ & $2.8\times10^{45}$ &\\\hline

\end{tabular}}
\vspace{0.2cm}
\label{Table-6}
\end{table*}

\begin{table*}
\centering
\caption{\textbf{Parameters used for modelling multi-wavelength data from extended jet of PKS 1354+195}}

\scalebox{1.0}{
\begin{tabular}{|c|c|c|c|c|c|c|c|c|c|} 

\hline

Knots & Dist(arcsec) & $\Gamma$ & $\theta_{obs}$ & $\delta$ & $B(mG)$ & Emission & $E_{min}(eV)$ & $E_{max}(eV)$ & $\alpha$\\ \hline

$A$ & 1.7 & 2.7 & $20^{\circ}$ & 6.14 & 0.2 & $1st\ population\ e^{-}synch$ & $10^7$ & $2.81\times10^{11}$ & 2.4\\ \cline{7-10}
&&&&&& $2nd\ population\ e^{-}synch$ & $5.62\times10^{12}$ & $3.16\times10^{13}$ & 2.3\\ \hline

$B$ & 4 & 2.8 & $20^{\circ}$ & 5.92 & 0.1 & $1st\ population\ e^{-}synch$ & $10^7$ & $2.51\times10^{11}$ & 2.4\\ \cline{7-10}
&&&&&& $2nd\ population\ e^{-}synch$ & $5.62\times10^{12}$ & $3.16\times10^{13}$ & 2.3\\ \hline

\end{tabular}}
\vspace{0.2cm}
\label{Table-7}
\end{table*}

\begin{table*}
\centering
\caption{\textbf{Magnetic field energy densities of different knots and total jet power for object PKS 1354+195}}

\scalebox{1.0}{
\begin{tabular}{|c|c|c|c|c|}\hline

Knots & $u'_{B}(erg/cm^3)$ & {$R$} {cm} & $P_{B}$(erg/sec) & $P_{total}$(erg/sec)\\ \hline

$A$ & $1.59\times10^{-9}$ & $3.086\times10^{21}$ & $1.04\times10^{46}$ & $1.32\times10^{46}$\\\cline{1-4}

$B$ & $3.98\times10^{-10}$ & $3.086\times10^{21}$ & $2.8\times10^{45}$ &\\\hline

\end{tabular}}
\vspace{0.2cm}
\label{Table-8}
\end{table*}

\begin{table*}
\centering
\caption{\textbf{Parameters used for modelling multi-wavelength data from extended jet of PKS 2209+080}}

\scalebox{1.0}{
\begin{tabular}{|c|c|c|c|c|c|c|c|c|c|} 

\hline

Knots & Dist(arcsec) & $\Gamma$ & $\theta_{obs}$ & $\delta$ & $B(mG)$ & Emission & $E_{min}(eV)$ & $E_{max}(eV)$ & $\alpha$\\ \hline

$A$& 0.52& 2.8 & $20^{\circ}$ & 5.92 & 0.3 & $1st\ population\ e^{-}synch$ & $10^7$ & $3.16\times10^{10}$ & 2.3 \\ \cline{7-10}
&&&&&& $2nd\ population\ e^{-}synch$ & $10^{11}$ & $3.16\times10^{12}$ & 2.3\\ \hline

$B$ &1.3 & 2.7 & $20^{\circ}$ & 6.14 & 0.2 & $1st\ population\ e^{-}synch$ & $10^7$ & $1.58\times10^{11}$  & 2.1\\ \cline{7-10}
&&&&&& $2nd\ population\ e^{-}synch$ & $3.98\times10^{11}$ & $10^{13}$  & 2.3 \\ \hline

$C$ & 2 & 2.6 & $20^{\circ}$ & 6.37& 0.1 & $1st\ population\ e^{-}synch$ & $10^7$ & $1.51\times10^{11}$ & 2.2\\ \cline{7-10}
&&&&&& $2nd\ population\ e^{-}synch$ & $7.94\times10^{11}$ & $10^{13}$  & 2.3 \\ \hline

$D$ & 3.2& 2.5 & $20^{\circ}$& 6.63 & 0.08 & $1st\ population\ e^{-}synch$ & $10^7$ & $1.34\times10^{11}$ &  2.1\\ \cline{7-10}
&&&&&& $2nd\ population\ e^{-}synch$ & $10^{13}$ & $3.16\times10^{13}$ & 2.3\\ \hline

$E$ & 4.7 & 2.4 & $20^{\circ}$ & 6.9 & 0.05 & $1st\ population\ e^{-}synch$ & $10^7$ & $2\times10^{11}$ & 2.3\\ \cline{7-10}
&&&&&& $2nd\ population\ e^{-}synch$ & $10^{12}$ & $3.16\times10^{13}$ & 2.3\\ \hline

\end{tabular}}
\vspace{0.2cm}
\label{Table-9}
\end{table*}

\begin{table*}
\centering
\caption{\textbf{Magnetic field energy densities of different knots and total jet power for object PKS 2209+080}}

\scalebox{1.0}{
\begin{tabular}{|c|c|c|c|c|}\hline

Knots & $u'_{B}(erg/cm^3)$ & {$R$} {cm} & $P_{B}$(erg/sec) & $P_{total}$(erg/sec)\\ \hline

$A$ & $15.92\times10^{-8}$ & $3.086\times10^{21}$ & $2.52\times10^{46}$ & $4\times10^{46}$\\\cline{1-4}

$B$ & $7.8\times10^{-8}$ & $3.086\times10^{21}$ & $1.04\times10^{46}$ &\\\cline{1-4}

$C$ & $10.83\times10^{-8}$ & $3.086\times10^{21}$ & $2.41\times10^{45}$ &\\\cline{1-4}

$D$ & $3.98\times10^{-8}$ & $3.086\times10^{21}$ & $1.4\times10^{45}$ &\\\cline{1-4}

$E$ & $12.89\times10^{-8}$ & $3.086\times10^{21}$ & $5.14\times10^{44}$ &\\ \hline

\end{tabular}}
\vspace{0.2cm}
\label{Table-10}
\end{table*}

\begin{table*}
\centering
\caption{\textbf{Parameters used for modelling multi-wavelength data from extended jet of PKS 0637-752}}

\scalebox{1.0}{
\begin{tabular}{|c|c|c|c|c|c|c|c|c|} 

\hline

Knots & $\Gamma$ & $\theta_{obs}$ & $\delta$ & $B(mG)$ & Emission & $E_{min}(eV)$ & $E_{max}(eV)$ & $\alpha$\\ \hline

$WK_{5.7}$ & 2.8 &$30^{\circ}$ & 5.92 & 0.1 & $1st\ population\ e^{-}synch$ & $10^{7}$ & $1.12\times10^{11}$ & 2.3\\ \cline{6-9}
&&&&& $2nd\ population\ e^{-}synch$ & $8.91\times10^{11}$ & $1.25\times10^{13}$ & 2.3\\ \hline

$WK_{7.8}$ & 2.7 & $30^{\circ}$ & 6.14 & 0.3 & $1st\ population e^{-}synch$ & $10^{7}$ & $3.16\times10^{11}$ & 2.6\\ \cline{6-9}
&&&&& $2nd\ population\ e^{-}synch$ & $5.01\times10^{12}$ & $3.16\times10^{13}$ & 2.3\\ \hline

$WK_{8.9}$ & 2.6 & $30^{\circ}$ & 6.37 & 0.5 & $1st\ population\ e^{-}synch$ & $10^{7}$ & $2.51\times10^{11}$ & 2.35\\ \cline{6-9}
&&&&& $2nd\ population\ e^{-}synch$ & $3.98\times10^{12}$ & $3.16\times10^{13}$ & 2.3\\ \hline

$WK_{9.7}$ & 2.5 & $30^{\circ}$ & 6.63 & 0.7 & $1st\ population\ e^{-}synch$ & $10^{7}$ & $2\times10^{11}$ & 2.5\\ \cline{6-9}
&&&&& $2nd\ population\ e^{-}synch$ & $5.01\times10^{12}$ & $3.16\times10^{13}$ & 2.3\\ \hline

\end{tabular}}
\vspace{0.2cm}
\label{Table-11}
\end{table*}

\begin{table*}
\centering
\caption{\textbf{Magnetic field energy densities of different knots and total jet power for object PKS 0637-752}}

\scalebox{1.0}{
\begin{tabular}{|c|c|c|c|c|}\hline

Knots & $u'_{B}(erg/cm^3)$ & {$R$} {cm} & $P_{B}$(erg/sec) & $P_{total}$(erg/sec)\\ \hline

$WK_{5.7}$ & $3.98\times10^{-10}$ & $3.086\times10^{21}$ & $2.8\times10^{45}$ & $1.189\times10^{47}$\\ \cline{1-4}

$WK_{7.8}$ & $3.582\times10^{-9}$ & $3.086\times10^{21}$ & $2.34\times10^{46}$ & \\ \cline{1-4}

$WK_{8.9}$ & $9.95\times10^{-9}$ & $3.086\times10^{21}$ & $6.51\times10^{46}$ &\\ \cline{1-4}

$WK_{9.7}$ & $1.95\times10^{-8}$ & $3.086\times10^{21}$ & $1.276\times10^{47}$ &\\ \hline

\end{tabular}}
\vspace{0.2cm}
\label{Table-12}
\end{table*}
\end{document}